\documentstyle[pre,aps]{revtex} 

\newcommand{\bdm}{\begin{displaymath}}
\newcommand{\edm}{\end{displaymath}}
\newcommand{\be}{\begin{equation}}
\newcommand{\ee}{\end{equation}}
\newcommand{\ba}{\begin{eqnarray}}
\newcommand{\ea}{\end{eqnarray}}
\newcommand{\bino}[2]{\left( \begin{array}{c} #1 \\ #2 \end{array} \right)}

\begin{document}
\draft
\title{Criticality in Simple Models of Evolution}
\author{Jan de Boer,
        A.D. Jackson, and 
        Tilo Wettig}
\address{ 
  Department of Physics, State University of New York at Stony Brook, 
  Stony Brook, New York 11794-3800, USA}
\date{\today}
\maketitle
\begin{abstract}

We consider two, apparently similar, models of biological evolution
which have been claimed to exhibit self-organized critical behaviour.
A careful reanalysis of these models, including several new analytic
results for one of them, suggests that they are qualitatively
different.  We demonstrate the limitations of the mean field
description of these systems.  We argue that a more precise definition
of self-organized criticality is desirable and establish several
criteria in this connection.

\end{abstract}


% uncomment next line for narrow pages a la PRL
%\narrowtext
%%%%%%%%%%%%%%%%%%%%%%%%%%%%%%%%%%%%%%%%%%%%%%%%%%%%%%%%%%%%%%%%%%%

\section{Introduction}
\label{section1}

The phenomenon of ``self-organized criticality'' (SOC) has recently
emerged as a topic of considerable interest with potential
applications ranging from the behaviour of sand piles and the
description of the growth of surfaces to generic descriptions of
biological evolution
\cite{bak1,bak2,bak3,interface,bak4,bak5,bak6,quake2}.  It would
appear that there is no general agreement on a suitable definition of
self-organized criticality.  (As an example, see the recent discussion
in \cite{let1,let2}.)  Equally important, there do not appear to be
universally accepted ``black box'' tests for its presence or absence
based solely on observables.  (Formal definitions based on
microscopic, structural analogies can be useful in model studies.
They would seem to have limited value in deciding whether any given
physical system is or is not self-organized critical.)  As a
consequence, systems with a wide range of characteristics have all
been designated as ``self-organized critical''.  The purpose of the
present work is two-fold.  We shall point out some striking observable
differences between two ``self-organized critical'' models which have
a remarkable structural similarity.  We will argue that these tests
suggest that one of the two models should {\em not\/} be regarded as
critical.  At the least, these differences will suggest that
definitions should be refined to permit a distinction between such
models.  Along the way, we shall present a number of new analytic
results.

We shall concentrate on two closely related and simple models
introduced by Bak and Sneppen to mimic biological evolution. They were
introduced in \cite{baksnep1,baksnep2} and further analyzed in
\cite{bla}.  The models (as we shall use them) involve a
one-dimensional array of $N$ sites.  Initially, each site is assigned
a ``barrier'' drawn at random on the interval $0 \le b_i \le 1$.  At
each update of the model, one identifies the minimum barrier and
assigns a new random barrier to this minimum site.  In addition, we
assign new random barriers to a certain number of other sites. (The
new barriers are again drawn uniformly on the interval $[0,1]$.)  In
the first version of the model, we choose the two nearest neighbours
of the minimum (the local or nearest neighbour model).\footnote{The
biological interpretation of this model is simple: Each site
represents a species.  The location of each species represents its
order in a ``food chain''.  The barrier assigned to each species is a
measure of its ``survivability''.  At each update, the least
survivable species undergoes a mutation and induces mutations in
random sites of its nearest neighbours in the food chain.} In the
second version, new barriers are assigned to $K-1$ additional sites
which are chosen at random (the random neighbour model).  We adopt
periodic boundary conditions in each model.

As we shall indicate, the nearest neighbour model is the richer of the
two.\footnote{We believe that this model clearly displays
self-organized criticality (in the limit $N \rightarrow \infty$).} It
has a number of fascinating properties.  The most important of these
is that it displays ``punctuated equilibrium'' \cite{punceq1,punceq2}
in which any given (small) segment of the sites will experience long
periods of inactivity punctuated by brief periods of violent activity.
(Hence, its interest as a model of evolution.)  In equilibrium, the
nearest neighbour model has a non-trivial distribution of barriers
\cite{baksnep1}.  One finds that virtually all of the barriers are
uniformly distributed in the interval $2/3 \le b_i \le 1$.  In the
thermodynamic limit, $N \rightarrow \infty$, only a vanishing fraction
of barriers are in the region $0 \le b_i \le 2/3$.  There is also a
non-trivial distribution of the minimum barriers selected at each
update.  The probability of finding a minimum barrier with the value
$b$ decreases approximately linearly\footnote{One can show
analytically that this probability distribution is not exactly a
linear function even in the limit $N\rightarrow\infty$. It is,
however, very close to linear.  In this regard, see Fig.~2 of
Ref.~\cite{baksnep1}.} to zero as $b$ increases from $0$ to $2/3$.
The fraction of minimum barriers having values greater than $2/3$ also
vanishes in the thermodynamic limit.

One can define an ``avalanche'' in this model by first establishing a
threshold ($\lambda < 2/3$) and, at each update, counting the total
number of ``active sites'' with barriers less than $\lambda$.  An
avalanche of temporal extent $T$ is said to occur when there are
active sites for precisely $T$ consecutive updates of the model.  In
the limit $\lambda \rightarrow 2/3$, one finds (numerically) a power
law distribution of avalanches.  Specifically, $P_{\rm aval}(T) \sim
T^{-1.08}$ for large $T$ \cite{Paczuski}.  This power law provides the
first suggestion of critical behaviour in the model.

One can also study a variety of correlation functions (following
\cite{interface}) associated with the space or time movement of the
minimum barrier.  For example, the probability that the minimum
barriers at two successive updates will be separated by $x$ sites is
given by a power law, $P_x (x) \sim x^{-3.1}$ for large
$x$\cite{baksnep1}.  The most natural temporal correlation to study is
the probability that, if a given site was the minimum at some time, it
will also be the minimum at time $t$ later.  For large $t$, this
correlation is also given by a power law, $P_t (t) \sim t^{-0.42}$
\cite{snepjens}. Power-law correlations in {\em both} space and time
would seem to be the most compelling indicators of the presence of a
critical phenomenon.  We argue below that power laws in temporal
correlation functions alone are probably not a sufficient criterion
for criticality and that subsidiary conditions on the exponents are
needed.

To make this point, consider the simpler random neighbour variant of
the model \cite{baksnep2} in the thermodynamic limit.  This model also
has a non-trivial distribution of barriers.  In this case, virtually
all barriers are uniformly distributed over the interval $1/K \le b_i
\le 1$.  The fraction of barriers in the region $0 \le b_i \le 1/K$
vanishes.  There is again a non-trivial distribution of the minimum
barriers selected at each update.  The probability of finding a
minimum barrier with the value $b$ is constant over the region $0 \le
b \le 1/K$.  Only a vanishing fraction of the minimum barriers have
values greater than $1/K$. It can be shown analytically\footnote{The
case $K=2$ has been discussed in \cite{bla}, and the general proof is
given here in Appendix~\ref{appendix2}.} that avalanches again follow
a power law distribution (provided that one chooses the threshold to
be $\lambda=1/K$) with $P_{\rm aval}(T) \sim T^{-3/2}$.

The spatial correlations of the random neighbour model are completely
different from those found in the original local neighbour model.  In
the random neigbour model, there is a significant probability that the
minimum barrier will be at the same site at two successive updates.
As could be anticipated from the fact that there is no notion of
spatial ordering in the model to begin with, there are no other
spatial correlations in this model.  Spatial correlations have been
degraded from power laws to a delta function.  However, the temporal
correlation of the minimum (as defined above) is still given by a
power law, $P_t(t)\sim t^{-3/2}$ for large $t$ (see
Sec.~\ref{section4}).  We will argue in Sec.~\ref{section4} that the
existence of this power law alone is not a good indicator of
criticality.  The value of the exponent is crucial.  We will argue
that the normalizability of $P_t(t)$ provides a useful indication that
the system is not critical.  By this criterion, we would conclude that
the random neighbour variant does not describe a critical system.  The
local neighbour variant does.

The original motivation for the random neighbour model was precisely
that it was easy to solve.  In particular, it allows for a
straightforward and relatively successful mean field description of
the barrier distribution \cite{baksnep2}.  However, the simplification
process results in the degradation of those power-law correlations
which provide the strictest test of criticality.  In fact, this mean
field description is not particularly successful at describing the
most interesting features of the model which are related to the
distribution of barriers.  The most interesting ``observable
quantity'' to be extracted from the barrier distribution is the
avalanche probability distribution. For this we need the distribution
of ``active'' sites with barriers below $1/K$ which (by definition)
contribute to avalanches.  These barriers comprise only a vanishing
fraction of of all barriers in the thermodynamic limit.  Further, all
avalanches must end.  (Otherwise, there would be little merit in
talking about avalanche distributions.)  Thus, the fluctuations in the
number of active barriers must be comparable to its average value.
This contradicts the usual assumption of mean field theory that
fluctuations of quantities are small compared to their average value.
Under such circumstances, the nature of a mean field description
renders it of limited utility.  It is good at describing the features
of the critical state of the system.  It is not suitable for providing
a description of the interesting fluctuations about this critical
state.  On the other hand, the number of sites with barriers larger
than $1/K$ is ${\cal O}(N)$.  In this case, fluctuations about the
mean field results are genuinely small (i.e., of ${\cal
O}(\sqrt{N})$), and their mean field description can be (and is)
successful.  Unfortunately, such sites are largely inert and of little
physical interest.

The organization of this paper is as follows.  In Sec.~\ref{section2}
we review the mean field approximation to the random neighbour model
\cite{baksnep2} and an exact analysis of this model in terms of the
number of active sites \cite{bla}. We show how to obtain the
distribution of barrier heights exactly and contrast the results with
the approximate values obtained with mean field theory.  In
Sec.~\ref{section3} we shall indulge in a brief digression in order to
demonstrate that the power laws in ``avalanche'' distributions are not
necessarily indicative of critical behaviour.  We shall make this
point by considering a one-dimensional random walk and an equivalent
one-dimensional chain of non-interacting spins.  (We take the position
that these systems should not be regarded as critical.  Others may
disagree.)  In Sec.~\ref{section4} we consider the role of spatial and
temporal correlations in the random and local neighbour models.  Here,
as noted above, we will encounter a significant difference between the
random and local neighbour models.  A number of conclusions will be
drawn in Sec.~\ref{section5}, along with suggestions for possible
further tests of criticality.  Two appendices are provided for a
detailed account of the derivation of various analytic results.

\section{Mean Field Theory and Beyond}
\label{section2}

In this section we wish to summarize the description of the random
neighbour model using mean field theory \cite{baksnep2}.  Since it is
our intention to assess the validity of mean field theory, we will
also introduce a simple and tractable reformulation of the model
\cite{bla} which leads to exact results without the need for
simulations.

In the mean field theory approach to the random neighbour model, we
replace the joint probability distribution $\rho(b_1,\ldots,b_N)$ of
the barrier heights by $p(b)^N$, where $p(b)$ is the average
equilibrium distribution of barrier heights at one site and $N$ is the
number of sites in the model. It is easiest to work with $\phi (b)$
defined as 
\be 
  \phi(b) = \int_b^1 db' \ p(b') \ \ .
\label{phi_of_b}
\ee
Evidently, $p(b)$ follows immediately by differentiation.  Further,
the probability distribution $C(b)$ for the value of the minimum
barrier is also determined by $\phi (b)$, 
\be 
  C(b) = - \frac{d\phi^N}{db} \ \ .  
\label{C_of_b}
\ee 
By demanding equilibrium one can derive the following simple equation 
for $\phi (b)$, 
\be 
  (N-K) \phi^N + N(K-1) \phi + K (N-1)(b-1) = 0 \ \ .  
\label{meanfield}
\ee 
Here, we have allowed $K$, the total number of barriers changed at 
each update, to assume any value $K \ge 1$. 

Now, let us rewrite the random neighbour model by concentrating on the
number of active sites, $A$, at any given update \cite{bla}.  An
active site will be any site which has a barrier height less than some
threshold value $0 \le \lambda \le 1$.  The number of active sites can
change at any update of the model.  The probability for a given change
will depend on the current number of active sites.  A detailed
analysis of this active site description is performed in
Appendix~\ref{appendix1} (for finite $N$) and in
Appendix~\ref{appendix2} (in the thermodynamic limit $N \rightarrow
\infty$).

To obtain a sense of the nature of this model, it is
useful to consider the average number of active sites as a function of
the threshold $\lambda$.  To do this, consider the average change in
$A$ in one update if $A \ge 1$.  The calculation is done in
Appendix~\ref{appendix1}, and we obtain [see Eq.~(\ref{DeltaA})]
\be
  {\overline{\Delta A}} = (K\lambda-1) - \frac{(A-1)(K-1)}{N-1} \ \ .
\ee
First, consider the case $\lambda < 1/K$.  The value of
${\overline{\Delta A}}$ is less than $0$ for all values of $A$ and,
more importantly, for all values of $N$.  As we approach the
thermodynamic limit $N \rightarrow \infty$, we see that
${\overline{\Delta A}}$ will approach some limiting negative number
with additional negative corrections of ${\cal O}(1/N)$.  It is clear
that the average value of ${\overline{\Delta A}}$ will be positive
when $A$ is precisely zero.  Thus, for $\lambda < 1/K$, the average
number of active sites has some value, ${\overline{A}}(\lambda)$,
which evidently depends on $\lambda$ but is independent of $N$.  The
density of active sites, defined as $\rho = {\overline{A}}/N$, is
precisely $0$ in the thermodynamic limit for all $\lambda < 1/K$.

For $\lambda > 1/K$ the situation is quite different.  Now
${\overline{\Delta A}}$ is {\em positive\/} for all values of $A$ less
than some $N x_c (\lambda)$ and negative for all $A$ greater than $N
x_c (\lambda)$.  The value of $x_c (\lambda)$ is readily seen to be
\be
  x_c (\lambda) = \frac{K\lambda-1}{K-1} + \frac{1}{N}
                  \frac{K(1-\lambda)}{K-1} \ \ .
\label{xcrit}
\ee
In this case, the average number of active sites will grow with $N$.
The density of active sites will be $x_c (\lambda)$.  In other words,
in the thermodynamic limit,
\be
\rho = \left\{ 
         \begin{array}{cl}
           0 & {\rm for} \ \ \lambda < 1/K \\
           (K\lambda-1)/(K-1)  & {\rm for} \ \ \lambda > 1/K 
         \end{array}
       \right. \ \ .
\label{density}
\ee
The density of active sites grows linearly from $0$ to $1$ as
$\lambda$ grows from $1/K$ to $1$.  At the point $\lambda = 1/K$, the
density is continuous but its derivative is discontinuous.  This
result is superficially suggestive of a second-order transition with a
critical point at $\lambda = 1/K$.

Let us now determine the average number of active sites for $\lambda <
1/K$ in the thermodynamic limit.\footnote{This quantity follows
immediately from Eq.~(\ref{density}) when $\lambda > 1/K$.} In
Appendix~\ref{appendix2}, we have derived the generating function for
the probability of finding precisely $A$ active sites,
$\Omega(z)=\sum_{A\ge 0} P(A)z^A$. One finds
\begin{equation}
  \Omega(z)=(1-f'(1))\frac{(1-z)f(z)}{f(z)-z} \ \ ,
\end{equation}
where
\begin{equation}
  f(z)=(1-\lambda+\lambda z)^K \ \ .
\end{equation}
From this, the explicit values of $P(A)$ can be extracted by means of
contour integration \cite{bla}.  The average value of the number of
active sites is
\begin{equation}
  {\overline{A}} = \sum_{A\ge 0} P(A) A =\Omega'(1)= 
  K\lambda \frac{1- (K+1) \lambda /2}{1-K \lambda} \ \ .  
\label{abar}
\end{equation}
As noted, this result applies for $\lambda < 1/K$.  The
root-mean-square deviation from $\overline{A}$ can also be calculated
from $\Omega(z)$.  We find that
\ba
  \delta A & = & \left[\Omega''(1)+\Omega'(1)-(\Omega'(1))^2\right]^{1/2}
                 \nonumber \\
  & = & \frac{\sqrt{K\lambda}}{1-K\lambda} 
  \left[ 1- \frac{3}{2}(K+1)\lambda + \frac{1}{6}(K+1)(5K+4)\lambda^2
         -\frac{1}{12}K(K+1)(K+5)\lambda^3 \right]^{1/2} \ \ . 
\ea
The important point to note is that $\delta A$ is larger than
${\overline{A}}$ for all $ \lambda < 1/K$ and equals ${\overline{A}}$
at the limit.

The result of Eq.~(\ref{abar}) is interesting.  It allows us to see
short-comings of the mean field analysis of this model for $\lambda <
1/K$.  In the mean field approach,
\begin{equation}
  {\overline{A}}_{\rm mf} (\lambda) = N [1 - \phi (\lambda)] \ \ .
\end{equation}
It is easy to solve the mean field approximation in the limit of
large $N$ and for $\lambda < 1/K$.  We find
\begin{equation}
  (1-K/N)(1-{\overline{A}}_{\rm mf}/N)^N + (K-1)(1-
  {\overline{A}}_{\rm mf}/N)+ K (1-1/N)(\lambda-1) = 0 
\end{equation}
which leads to 
\begin{equation} 
  {\overline{A}}_{\rm mf} = - \ln {(1 - K\lambda)} 
\label{abarmf}
\end{equation}
in the thermodynamic limit.  While Eqs.~(\ref{abarmf}) and
(\ref{abar}) agree in the limit $\lambda \rightarrow 0$, they differ
everywhere else.  These differences are particularly striking as
$\lambda$ approaches the critical value of $1/K$.

To obtain a quantitative sense for the differences between mean field
theory and the active site model, we compute the equilibrium
distributions of barriers, $p(\lambda)$, and of the minimum,
$C(\lambda)$, in each approach for finite $N$ and compare the results
with simulations.  In the mean field approximation, we have to solve
Eq.~(\ref{meanfield}) numerically for $\phi(\lambda)$.  Using
Eqs.~(\ref{phi_of_b})--(\ref{meanfield}), we can derive explicit
expressions for $p(\lambda)$ and $C(\lambda)$ in terms of
$\phi(\lambda)$.  In the active site version, $p(\lambda)$ is obtained
from
\be
  p(\lambda)=\frac{1}{N}\frac{d\overline{A}}{d\lambda} \ \ .
\ee
Note that this definition agrees with that of mean field theory
through Eq.~(\ref{phi_of_b}).  Since ${\overline{A}} = \sum_{A\ge 0}
P(A) A$, we obtain
\be
  p(\lambda)=\frac{1}{N}\sum_{A\ge 0} P'(A) A \ \ ,
\label{pb}
\ee
where $P'(A)$ denotes the derivative of $P(A)$ with respect to
$\lambda$. The probability distribution of the minimum is simply
\be
  C(\lambda)=-P'(0) 
\label{Cb}
\ee
in the active site description.  In Appendix~\ref{appendix1} we
describe how to obtain $P(A)$ and $P'(A)$ for finite $N$.  Numerical
results for $p(\lambda)$ and $C(\lambda)$ obtained from mean field
theory, the active site analysis, and simulations for $K=3$ and
$N=100$ are displayed in Table~\ref{table1}.  The barrier and minimum
distributions for the active site and mean field calculations are
again different.  As expected, the (exact) active site results are in
agreement with the results of simulations.  The barrier distribution
in the region $\lambda < 1/K$ is, by nature, small. Thus, absolute
differences between mean field and active site results are also small.
However, this region contains the active sites which carry the
interesting physics.  The fractional difference between mean field and
active site results is large, and we regard this disagreement as
significant.

Of course, the mean field results are sometimes completely reliable.
In the thermodynamic limit and for $\lambda > 1/K$, we can solve the
mean field equations by noting that $\phi$ is now genuinely less than
$1$.  This justifies dropping the term $\phi^N$ to obtain exactly the
result of Eq.~(\ref{xcrit}) obtained with the active site calculation.
This success is not surprising.  In this domain, ${\overline{A}}$
grows linearly with $N$ while $\delta {\overline{A}}$ grows like
$\sqrt{N}$.

It is easy to understand these results.  The success of a mean field
calculation of any quantity requires that the ratio of the rms
deviation of this quantity to its average value vanishes in the
thermodynamic limit.  When the deviation of a quantity is of the same
order as its average value, the mean field approximation is likely to
fail.  Indeed, the fact that $\delta A \ge \overline{A}$ for all
$\lambda \le 1/K$ is a clear indicator that mean field approximations
are not likely to be reliable in this region.  Since one anticipates
large fluctuations to be the hallmark of any critical system, there is
virtually an {\em a priori\/} contradiction in expecting that mean
field theory will provide a complete description of any critical
system.

\section{Power Laws without Criticality}
\label{section3}

Some authors have used the existence of power laws in the distribution
of avalanches as a function of their duration as an indicator of
criticality in similar systems.  We have already mentioned in
Sec.~\ref{section1} that both the local and the random neighbour model
show such power laws with exponents $-1.08$ (for the local version
with $\lambda=2/3$) and $-3/2$ (for the random version with
$\lambda=1/K$).  One notes the existence of avalanches ``of all size
scales'' (meaning no exponentials) and declares criticality.  It is
thus useful to consider a system which is non-critical by common
consent and which nevertheless has such power laws --- a random walk
in one dimension.

In this case, we define the start of an ``avalanche'' as being when
the walker is at the point $x=0$.  Now consider a ($P$-type) walk for
which the walk ends as soon as $x$ is again 0.  We are interested in
$P_{2n}$ which is the probability that a walk will return to $x=0$
(for the first and only time) at step $2n$.  To determine this
probability, it is useful to consider a $Q$-type walk of length $2n$
for which $x=0$ at the end points without regard for whether $x=0$ is
also reached at intermediate steps.  This walk occurs with probability
$Q_{2n}$.  Evidently, $Q_{2n}$ is trivial to determine from
combinatorics:
\begin{equation}
  Q_{2n} = \frac{1}{2^{2n}} \bino{2n}{n} \ \ .
\label{Q2n}
\end{equation}
Any $Q$-type walk is either a pure $P$-type walk or can be decomposed
into some sequence of $P$-type walks of varying lengths provided only
that the sum of their lengths is $2n$. Using this observation, we
deduce that the generating function for the numbers $Q_{2n}$ is given
by
\begin{equation}
  1+\sum_{n=1}^{\infty} Q_{2n} z^{2n}=\frac{1}{1-\sum_{n=1}^{\infty} 
  P_{2n} z^{2n}} \ \ .
\label{Q_from_P}
\end{equation}
Summing over all walks has led to great simplification.  The various
terms contributing to the right hand side at a given order merely
describe the various ways in which a $Q$-walk can be built from
$P$-walks.

Fortunately, the left side of Eq.~(\ref{Q_from_P}) can be summed
analytically given the form of Eq.~(\ref{Q2n}).  Thus,
\begin{equation}
  1 - \sum_{n=1}^{\infty} P_{2n} z^{2n} = \sqrt{1 - z^2} \ \ .
\end{equation}
This allows us to write
\begin{equation}
  P_{2n} = \frac{(2n-3)!!}{(2n)!!} \rightarrow 
  \sqrt{\frac{2}{\pi}} \frac{1}{(2n)^{3/2}} 
\end{equation}
where the last expression is the large $n$ limit.  

It is no accident that an identical power is found for the avalanche
distribution of the random neighbour model.  The results of the last
section (and of the next) indicate that this distribution function
also comes from a random walk of a very similar character.  A $t$-step
avalanche starts when the number of active sites changes from zero to
non-zero and ends when the number of active sites is again 0 at step
$t$.  The rules for changing the number of active sites, given in
Appendix~\ref{appendix2}, are simply a set of random walk rules.  The
probability for an avalanche of length $t$ (with $\lambda=1/K$) in the
random neighbour model is found from Eq.~(\ref{B15}) to be
\be 
  P_{\rm aval}(t) = \sqrt{\frac{K}{2\pi (K-1)}} t^{-3/2} 
                    + {\cal O}(t^{-5/2}) \ \ .
\ee 
This exponent is identical to the random walk exponent because the
problems are essentially identical. This is only true for
$\lambda=1/K$ since then the average change $\overline{\Delta A}$ of
Eq.~(\ref{DeltaA}) equals zero (in the thermodynamic limit) so that we
are dealing with an unbiased random walk.

To underscore the fact that this particular power law is not a good
indicator of criticality, consider a one-dimensional chain of spins
$\pm 1$ which do not interact.\footnote{This system is not critical
even when there is an interaction.  Everything which follows can be
generalized to include an arbitrary nearest-neighbour interaction and
a magnetic field of arbitrary strength using text book techniques.} As
usual, we adopt periodic boundary conditions.  For simplicity, we
consider only the thermodynamic limit as the number of sites goes to
infinity.  The only question is how to define an avalanche.  We choose
to consider the distribution of the sizes of domains of zero
magnetization.  In other words, start at any site and count the number
of sites until the total magnetization is zero.  The first zero
encountered defines the size of the domain.  The ensemble average
distribution of domain sizes, $P(t)$, is precisely given by the random
walk problem solved above.  Thus, $P(t) \sim t^{-3/2}$.  Clearly, this
power law does not indicate that the non-interacting spin lattice is
critical.

Since there is at least one example of a manifestly non-critical
system which exhibits a power law in an avalanche distribution, the
presence of such power laws cannot be regarded as a sufficient
indicator of critical behaviour.

\section{The Role of Correlations}
\label{section4}

So far, the behaviours of the local and random neighbour models are
qualitatively identical.  They each have a non-trivial distribution of
barrier heights and a corresponding distribution of the heights of
minimum barriers.  Each has a power law avalanche distribution.  It is
our {\em a priori\/} expectation that the local neighbour model is a
genuine self-organized critical model.  The fact that the random
neighbour variant is essentially identical to a random walk and,
hence, to a non-interacting one-dimensional spin chain strongly
suggests that this model is not critical.  We would thus like to find
a test which distinguishes between these models.  It would seem
natural to consider the spatial and temporal correlation between
localized avalanches.  Since a sound definition of local avalanches is
somewhat subtle, we consider instead the spatial and temporal
correlations between the {\em minimum\/} barrier (following
\cite{interface}).  This has the virtue that it is very much like the
kind of question one would ask to determine criticality in condensed
matter systems or in statistical mechanics.  It is also a sensible
first step.  For the local neighbour model, all sites in a single,
local avalanche must have been activated during the avalanche.  For
every site in the local avalanche, the minimum barrier must have
passed through the site itself or through one of its nearest
neighbours.  Hence, the path of the minimum provides the ``skeleton''
upon which the full avalanche will be built.  Another virtue of
working with the minimum barrier is that there is neither need nor
room for the somewhat artificial introduction of a threshold or any
special definition of an active site.

In the nearest neighbour model, results for the spatial and the
temporal correlation functions can currently only be obtained by
simulations.  The spatial correlation function $P_x(x)$ is defined as
the probability that the minimum barriers at two successive updates
are separated by $x$ sites.  There are two temporal correlation
functions of interest.  The first-return probability, $S(t)$, is
defined as the probability that, if a given site is the minimum at
time $t_0$, it will again be the minimum {\em for the first time\/} at
time $t_0+t$. The all-return probability, $P_t (t)$, is the
probability that this site will also be the minimum at $t_0 +t$
regardless of what happens at intermediate times.  In all cases, we
observe power law behaviour for large arguments.  Specifically, for
large $x$ we have \cite{baksnep1}
\be
  P_x (x) \sim \frac{1}{x^{3.1}} \ \ .
\ee
For large $t$ the first-return probability is found to be 
\be
  S (t) \sim \frac{1}{t^{1.6}} \ \ .
\ee
The all-return probability for large $t$ is \cite{snepjens}
\be
  P_t (t) \sim \frac{1}{t^{0.42}} \ \ .
\label{Ptlocal}
\ee
Numerical results for the spatial correlation function and the
temporal correlation function (first- and all-returns) were obtained
by simulations and are shown in Fig.~\ref{figPxlocal} and
Fig.~\ref{figPtlocal}, respectively.  The spatial correlation function
is familiar from Bak and Sneppen \cite{baksnep1} who describe the
motion of the minimum as a random walk in which the distance moved by
the minimum goes like $t^{1/3}$.  For any number of sites $N$, the
all-returns temporal correlation function follows the power law of
Eq.~(\ref{Ptlocal}) until it hits the accidental coincidence rate of
$1/N$.  It does this when
\begin{equation}  
  \Delta t \approx 0.05\, N^{2.4} \ \ .
\end{equation}
The interpretation of this result is simple.  The power law
correlations do not extend over times longer than the longest (local)
avalanche in the system.  The spatial extent of this longest avalanche
will be on the order of $N$.  The above equation tells us roughly the
relation between the largest (spatial) avalanche and its (temporal)
duration.  Note that the exponent $1/2.4 = 0.42$ is {\em not\/} equal
to the value of $1/3$ which one might crudely have expected from Bak
and Sneppen.  While it remains to be seen, we might expect that the
duration of an avalanche of spatial size $n$ will be of order
$n^{2.4}$.

Let us now consider the random neighbour model.  Here, the
correlations can be obtained analytically.  First, consider the
spatial correlation between the minimum at successive updates.  There
is some probability that the minimum will stay at the same site. In
the thermodynamic limit, this probability is given by $p_1=1-(K-1)\ln
[K/(K-1)]$, see Eq.~(\ref{aux3}).  Otherwise, the distribution of next
minima is random (i.e., spatially uniform).  Results for this
correlation function have also been obtained by simulation and are
shown in Fig.~\ref{figPxrandom}.  As expected, there is complete
agreement between the analytic results and the simulation.

Analytical results for the temporal correlation function are more
difficult to obtain.  An exact calculation of the first-return
probabilities, $S(t)$, is given in Appendix~\ref{appendix2} in the
thermodynamic limit.  Similar calculations for finite $N$ are possible
but tedious.  The leading term in $S(t)$ in the thermodynamic limit is
\be
S(t) = \frac{1}{3K} \sqrt{\frac{2(K-1)}{\pi K}} t^{-3/2} 
          + {\cal O}(t^{-5/2}) \ \ .
\ee
Additional terms in the asymptotic expansion of $S(t)$ are given in
Appendix~\ref{appendix2}.  Convergence of $S(t)$ for finite $N$ to the
values given here in the thermodynamic limit is remarkably slow.
Thus, Fig.~\ref{figPtrandom} shows a comparison of these analytic
results with the first-return probabilities arising from simulations
for the case $N = 10^5$.  While the agreement is extremely good, it is
not perfect.  Since the arguments in Appendix~\ref{appendix2} are
exact, several comments are in order.  First, the small discrepancies
apparent for small $t$ are not due to finite $N$ effects.  Rather,
they are due to our use of (three) leading terms an the asymptotic
expansion of $S(t)$.  In spite of this truncation, agreement is
excellent.  The discrepancies at large $t$ are finite $N$ effects.

Analytic calculation of the all-return probability, $P_t (t)$, is more
challenging.  The na\"{\i}ve expectation is that all-return
probabilities can be determined using an equation similar to
Eq.~(\ref{Q_from_P}), i.e.,
\be
  1+\sum_{n=1}^{\infty} P_t (n) z^n \stackrel{?}{=}
       \frac{1}{1-\sum_{n=1}^{\infty}S(n) z^n} \ \ .
\label{approx_rel}
\ee
In fact, this equation is not exact since it involves ensemble
averages at an inappropriately early stage in the calculation (see
Appendix~\ref{appendix2} for details).  It is far from trivial to
solve for $P_t(t)$ exactly due to reasons which will be discussed in
Appendix~\ref{appendix2}. There, we derive an improved equation for
$P_t(t)$. However, an explicit analytic solution of this equation is,
at best, tedious. Therefore, we have elected to perform an approximate
calculation using Eq.~(\ref{approx_rel}).  We can use the results of
simulations to test the approximate validity of Eq.~(\ref{approx_rel})
in relating all-returns to first-returns.  It is surprisingly accurate
for both the local and random neighbour models and results in errors
which are, at worst, at the 1\% level.  The result for large $t$ is
given in Eq.~(\ref{allreturn}) in the thermodynamic limit.  We find
\be 
  P_t(t) = \frac{K}{3(K-1)^2}\sqrt{\frac{2(K-1)}{\pi K}}t^{-3/2} +
           {\cal O}(t^{-5/2}) \ \ .
\label{Ptrandom}
\ee
For completeness, we have also performed simulations for this
quantity.  Again, convergence with $N$ to the thermodynamic limit is
extremely slow.  Thus, the results shown in Fig.~\ref{figPtrandom} are
for a simulation with $10^5$ sites.  The values of $P_t (t)$ assume
the expected $t^{-3/2}$ form quite quickly.  This form persists for
some time.  Since the number of sites is finite, there is a finite
probability (independent, of time) that a given site will
``accidentally'' be chosen to be updated.  For sufficiently large $t$,
this accidental rate of $1/N$ will dominate $P_t (t)$.  This fact is
clearly reflected in the form of Fig.~\ref{figPtrandom} for very large
times.  The discrepancies for small $t$ are larger than those observed
for the first-return probabilities.  This has nothing to do with
deviations from the thermodynamic limit or, worse, inadequacies in the
use of Eq.~(\ref{approx_rel}).  Rather, it is related to the fact that
the asymptotic expansion of $P_t (t)$ converges slowly.  This
discrepancy can be removed by noting that, as a consequence of
Eq.~(\ref{approx_rel}), $P_t (1) = S(1)$.  (This result is evidently
trivial and exact.)  The next few terms can be improved analogously by
expanding Eq.~(\ref{approx_rel}) rather than using the asymptotic
expansion Eq.~(\ref{Ptrandom}).  Except as noted in the asymptotic
region of large $t$, the discrepancies between our calculated values
of $P_t (t)$ and simulations are at the level expected from the
approximate validity of Eq.~(\ref{approx_rel}).

We are now in a position to assess the main qualitative differences
between the two versions of the model.  The local version displays
power law correlations in both space and time.  We view this as a
compelling indicator of criticality.  The random variant of the model
has no spatial correlations apart from a finite same-site correlation.
It would seem unfair to place too much emphasis on the absence of
spatial correlations in the random neighbour model since this was
clearly built in by design.  The random neighbour model does show
power law behaviour in the temporal correlation function, and it is
legitimate to discuss the differences in this correlation function for
the two versions of the model.

We believe that there are two reasons why the power laws of the random
neighbour model are not indicators of criticality.  The first argument
is of a qualitative nature.  The arguments presented in
Appendix~\ref{appendix2} leading to Eq.~(\ref{SDeltat}) rely solely on
the analogy to a random walk.  They require nothing more than the
ensemble averages of various one-body operators.  We believe that a
convincing demonstration of criticality must rely, either directly or
indirectly, on the demonstration of the existence of genuine
correlations.  Since the arguments leading to Eq.~(\ref{allreturn}) do
not require any information about two- or many-body operators, we do
not believe this power law (or any other power law obtained in a
similar fashion) should be regarded as a convincing indicator of
criticality.

The second argument is more quantitative.  While
Eq.~(\ref{approx_rel}) is not generally valid for either the random or
local models, we show in Appendix~\ref{appendix2} that it is true for
the random neighbour model in the special case $z=1$.  This allows us
to relate the sum over all-return probabilities to the sum over
first-return probabilities as
\be
  1+\sum_{n=1}^{\infty} P_t (n) = 
       \frac{1}{1-\sum_{n=1}^{\infty}S(n)} \ \ .
\label{inverse}
\ee
It is clear from the definition of the first-return probability that
the sum over all $S(n)$ must be finite and cannot exceed $1$
($\sum_{n\ge 1} S(n) = {\cal S} \le 1$).  By contrast, the sum over
all-return probabilities need not be finite since any given site can
be visited any number of times over an infinite time interval.  If
${\cal S}<1$, Eq.~(\ref{inverse}) indicates that the sum over the $P_t
(n)$ will be finite.  When this happens, it means that a given site
will be active only a finite number of times even though we study the
system for an infinite time interval.  It is not reasonable to regard
as critical any system in which individual sites ``die''.  On the
other hand, if ${\cal S}$ is exactly equal to $1$, the sum over the
$P_t (n)$ diverges.  In this case, sites will never die.  Thus, we
believe that a normalized sum over the $S(n)$ (${\cal S}=1$) and,
equivalently, a divergent sum over the $P_t (n)$ should be necessary
requirements for criticality.  (The equivalence of these criteria is
general and does not require the validity of Eq.~(\ref{approx_rel}).
If, upon reactivation, there is a finite probability that the site
will die, the sum over all-returns must be finite.  In practice, it is
usually easier to detect a divergence than it is to extract a
specific, finite normalization.)  Obviously, the condition for
convergence or divergence of the sum over the $P_t (n)$ can be
reformulated in terms of the exponent $\alpha$ in the asymptotic form
$P_t (n) \sim n^{\alpha}$.  If $\alpha<-1$, the model is not critical.

The random neighbour model does not satisfy these requirements.  We
see from Eq.~(\ref{aux5}) with $z=1$ that $\sum_{n\ge 1} S(n) =1/K<1$.
Thus, at any update when a given site becomes active, there is a
probability $1-1/K$ that it will never become the active site again.
The reason for this is clear from the known distribution of all
barriers and of minimum barriers: The distribution of all barriers
indicates that ${\cal O}(N)$ barriers have heights greater than $1/K$.
No barrier with height greater than $1/K$ will ever be the minimum
barrier (in the thermodynamic limit).  The new value drawn for the
barrier at the active site has a probability of $1 - 1/K$ of being
assigned a height greater than $1/K$.  When this happens, the site
dies and will never become the minimum again.  As a consequence, the
sum over $P_t (n)$ must converge.  In fact, $P_t (n) \sim n^{-3/2}$
for large $n$.  Hence, the random model does not satisfy this
criterion for criticality.  These considerations only apply in the
thermodynamic limit of infinitely many sites.  This provides no
limitations since it is only in this limit that the question of
criticality arises.

The situation is dramatically different in the local neighbour model.
Simulations reveal that $P_t (n)\sim n^{-0.42}$ for large $n$.  Thus,
the sum over the $P_t(n)$ must diverge, and the sum over first-return
probabilities must be identically $1$ in the thermodynamic limit.  The
local neighbour model passes this test of criticality.  On the
na\"{\i}ve basis of the one-body distributions of all barriers and
minimum barriers, this result is surprising.  A newly selected site
will draw a new barrier of height greater than $2/3$ with a
probability of $1/3$ and might be expected to die.  Of course, this
site can be re-activated through the presence of spatial correlations.
The value of this barrier can be reset to a value less than $2/3$
provided only that its nearest neighbour becomes the minimum.  The
mechanism for site re-activation provides the connection with our
previous point: One-body arguments cannot provide this re-activation,
and dynamical correlations are required.

Power law spatial correlations are a good indicator of criticality.
Power laws in temporal correlators are only useful indicators of
criticality when they meet the additional criteria noted above.  Given
this and the fact that these additional conditions on temporal
correlators are normally enforced through the agency of spatial
correlations, it might seem pointless to consider temporal
correlations at all.  This is not the case.  It is easy to devise
sensible models for which the extraction of information regarding
spatial correlation is either difficult or impossible.  For example,
consider a ``quenched'' version of the current model in which the
additional sites that are to be changed when a given site becomes the
minimum are chosen according to some complicated (but fixed) scheme
according to which we guarantee only that each site has the same
number of ``neighbours''.  (In particular, the designation of
``neighbour'' need not be reversible.)  In this case, the definition
of a spatial separation between sites may be subtle and may be
ambiguous.  However, the temporal correlation function for the minimum
can be studied as above even for such systems.  Temporal correlators
can permit decisions regarding criticality even in cases where spatial
correlation functions are unavailable.

\section{Discussion and Conclusions}
\label{section5}

We have considered a local neighbour model of evolution and a
simplified random neighbour variant.  The random model was originally
introduced to permit an approximate analytic description of the model
using mean field theory.  We have demonstrated that it is potentially
dangerous to modify critical models in order to facilitate mean field
calculations.  Mean field theory assumes that fluctuations are small
while criticality insists that they are large.  In practice, mean
field theory does not provide an adequate description of the
distribution of active sites even in the random neighbour model.  Such
disagreements are not merely quantitative.  Consider, for example, the
distribution of all barriers as $\lambda = 1/K - \delta \lambda$
approaches the threshold value of $1/K$.  For any finite value of
$\delta \lambda$ it is always possible to consider a value of $N$
sufficiently large that the results of Eqs.~(\ref{abar}) and
(\ref{abarmf}) (obtained in the thermodynamic limit) apply.  These
results indicate that $p_{\rm mf}(\lambda) \sim 1/\delta \lambda$.
The exact results indicate that $p(\lambda) \sim 1/\delta \lambda^2$.
The results of Table~\ref{table1} indicate the presence of similar
qualitative errors in the mean field description of the distribution
of minimum barriers.  Fortunately, as we have indicated here, most of
the interesting aspects of the random neighbour model permit exact
analytic solution.

Both local and random neighbour models have non-trivial distributions
for all barriers and minimum barriers.  Each possesses a power-law
avalanche distribution.  However, the $T^{-3/2}$ distribution
encountered in the random neighbour model is precisely analogous to
that of a one-dimensional random walk.  It does not strike us as a
reliable indicator of criticality.  The discussion in
Sec.~\ref{section3} suggests that it is never safe to draw conclusions
regarding criticality on the basis of avalanche distributions.

A more detailed consideration of spatial and temporal correlations
between minimum barrier locations reveals a clear distinction between
the two versions of the model.  The local neighbour model displays
power law correlations in space (with exponent $-3.1$) and time (with
exponent $-0.42$ for all-returns).  The random neighbour model lacks
the necessary power-law correlations in space (by design) and shows
power law correlations in time (with exponent $-3/2$ for all-returns).

It seems reasonable to extend the definition of self-organized
criticality in order to distinguish between these models.  Thus, we
think it useful to require simultaneous power law correlations in
space and time.  (Definitions of criticality which focus more closely
on correlations are closer in spirit to those adopted in statistical
and condensed matter physics.  This would appear to be a virtue.)
Moreover, we require that the temporal correlation function be
non-normalizable (i.e., it must not fall off faster than $1/t$). This
is equivalent to the requirement that sites must never ``die'' in the
thermodynamic limit.  The local neighbour model passes these two tests
while the random neighbour model fails.  This also suggests that any
simplified version of a critical model must, at the very least, retain
non-trivial correlations in its solution.  If all observables in a
model can be calculated from one-body operators, it is likely that
those features of the model responsible for criticality have been
approximated away.  This appears to be the case for the random
neighbour model considered here.

Power laws are ubiquitous in self-organized critical models.  They
appear in (i) the duration of ``global'' avalanches, (ii) the two-body
spatial correlation between minimum barrier sites, and (iii) the
two-body temporal correlation between minimum barrier sites.  There is
also a power law which relates the total number of active sites in a
``global'' avalanche to its duration, $T$.\footnote{The total number
of active sites in an avalanche is roughly the area of the avalanche
in a space-time plot.} It is probably safe to insist that all four of
these quantities must be power law if a model is to be considered
self-organized critical.  Power laws in all of these four quantities
are also a necessary condition for a space-time plot of the model to
be self-similar.  It is tempting to speculate that the best definition
of self-organized criticality may be the self-similarity of its
space-time plot.  Of course, self-similarity is a stronger statement
than the mere presence of power laws.  This question merits a closer
look.

%%%%%%%%%%%%%%%%%%%%%%%%%%%%%%%%%%%%%%%%%%%%%%%%%%%%%%%%%%%%%
\acknowledgments
We would like to thank B. Derrida and H. K. Flyvbjerg for useful
discussions.  
JdB was partially supported by the National Science Foundation
under grant no.\ PHY 93-09888.
TW and ADJ were partially supported by the US Department of Energy
under grant no.\ DE-FG02-88ER 40388.

%%%%%%%%%%%%%%%%%%%%%%%%%%%%%%%%%%%%%%%%%%%%%%%%%%%%%%%%%%%%%

\appendix
 
\section{Active Site Analysis for Finite $N$} 
\label{appendix1}

Here, we derive some analytical results for the active site
description of the random neighbour model for finite $N$.  At each
update the value of the barrier at the minimum site and at $K-1$ other
sites, selected at random, are replaced by new barriers which are
drawn at random on the interval $[0,1]$.

In general, we start from a configuration in which there are precisely
$A$ active sites with barrier heights less than some threshold value
$\lambda$.  We then consider how the number of active sites can change
following one update of the system.  These update rules result in a
description of the system as a random walk in $A$ with probabilities
for specific changes in $A$ which depend on the current value of $A$.
At each update, the net number of active sites can be reduced or
increased.  The first step in each update is to remove the $K$
barriers which are to be changed.  This can reduce the number of
active sites.  If $A=0$, no active sites can be removed. If $A > 0$,
we remove the minimum barrier (which must be an active site) and $r-1$
additional active sites (where $0\le r-1\le K-1$).  This reduces the
number of active sites by $r$. The associated probabilities,
$d_{r,A}$, for reducing the number of active sites by $r$ are given as
\cite{bla}
\be
  d_{r,A}=\frac{\bino{A-1}{r-1}\bino{N-A}{K-r}}{\bino{N-1}{K-1}} \ \ .
\label{drA}
\ee
One assigns $K$ new random barriers which can create a number of new
active sites, $t$, ranging from $0$ to $K$.  The probability for
creating $t$ active sites, $c_t$, is given by
\be
  c_t=\bino{K}{t}\lambda^t(1-\lambda)^{K-t} \ \ .
\label{ct}
\ee
The $c_t$ are just the coefficients of $z^t$ in the expansion of
$f(z)=(1-\lambda+\lambda z)^K$.  (We everywhere use the convention
that $\bino{n}{m}=0$ if $m<0$ or $m>n$.)  The fact that the location
of the $K-1$ additional sites is chosen at random implies that there
are no spatial correlations in the system (other than a finite
same-site correlation).  The coefficients $d_{r,A}$ and $c_t$ provide
complete information about the system.  In principle, knowledge of
their values (for all $\lambda$) is sufficient to calculate all
observables in this model.

Now, let $P(A)$ be the equilibrium probability of having $A$ active
sites and let $\Omega(z)=\sum_{A\ge 0} P(A) z^A$ be the corresponding
generating function.  The condition for equilibrium is obtained by
requiring that the $P(A)$ should not change when the system is
updated:
\ba
  \Omega(z) = \sum_{A\ge 0} P(A) z^A
  & = & P(0) f(z) + \sum_{A>0} P(A) z^A \sum_{r=1}^{K} d_{r,A} z^{-r}
        \sum_{t=0}^{K} c_t z^t \nonumber \\
  & = & f(z) \left[ \Omega(0) + \sum_{A>0} P(A) z^A g_A(z) \right] \ \ ,
\label{omegaN}
\ea
where $g_A(z)=\sum_{r=1}^K d_{r,A} z^{-r}$ for $A>0$.  Unfortunately,
$g$ depends on $A$, and it is not possible to derive a simple analytic
expression for $\Omega(z)$ except in the thermodynamic limit (see
Appendix~\ref{appendix2}).  Nevertheless, Eq.~(\ref{omegaN}) leads to
a set of $N$ homogeneous equations for the $P(A)$ which are soluble by
standard numerical techniques for any value of $N$ when supplemented
by the condition that the sum over all $P(A)$ equals $1$.

It is useful to note that a much faster numerical approach is
available in the special but important case when $\lambda < 1/K$.  In
this case, only a finite number of the $P(A)$ contribute (independent
of $N$), and the sums in Eq.~(\ref{omegaN}) can be truncated safely.
These equations can be solved by a simple iteration starting from,
e.g., $P(0)=1$.

Before discussing ways to solve Eq.~(\ref{omegaN}) exactly, it is
useful to compute an exact result needed in Sec.~\ref{section2}.  If
$A>0$, the probability for a change $\Delta A$ in the number of active
sites in one update of the model, $P_{\Delta A}$, is
\be
  P_{\Delta A}=\sum_{r=1}^K d_{r,A}c_{\Delta A+r} \ \ .
\ee
Assuming that $A>0$, the average change in $A$ in one update is
\ba
  {\overline{\Delta A}}
  & = & \sum_{\Delta A=-K}^{K-1} P_{\Delta A} \Delta A 
        = (f g_A)'(1) = K\lambda + g_A'(1) \nonumber \\ 
  & = & (K\lambda-1) - \frac{(A-1)(K-1)}{N-1} \ \ .
\label{DeltaA}
\ea

Let us now return to the solution of Eq.~(\ref{omegaN}).  As noted,
the most straightforward approach is to rewrite it as a matrix
equation for the $P(A)$ which can then be solved numerically.  In
order to compute the equilibrium distribution of barriers and of the
minimum we also need a matrix equation for the $P'(A)$ (defined as
$dP(A)/d\lambda$). Using the normalization condition $\sum_{A\ge 0}
P(A) =1$, we obtain for $A\ge 1$
\be
  P(A) = \left[ 1-\sum_{A'=1}^{N} P(A') \right] c_A + \sum_{r=1}^K
         \sum_{t=0}^K d_{r,A-t+r} c_t P(A-t+r) \ \ .
\label{P(A)}
\ee
This leads to the matrix equation
\be
  \sum_{A'=1}^N M_{AA'} P(A')=Y(A) \ \ ,
\ee 
where
\be
  M_{AA'} = \delta_{AA'} + c_A - \sum_{r=1}^K c_{A-A'+r} d_{r,A} 
\ee
and 
\be 
  Y(A) = c_A \ \ .
\ee
To obtain $p(\lambda)$ and $C(\lambda)$, we differentiate
Eq.~(\ref{P(A)}) with respect to $\lambda$.  Note that the only
additional $\lambda$-dependence lies in the $c_A$'s.  Using
$\sum_{A\ge 0} P'(A) =0$, we obtain for $A\ge 1$
\be
  P'(A) = - c_A \sum_{A'=1}^{N} P'(A') + c_A' P(0) + \sum_{r=1}^K
     \sum_{t=0}^K d_{r,A-t+r} [ c_t' P(A-t+r) + c_t P'(A-t+r) ] \ \ .
\ee
Here, $c_t'$ is simply
\be
  c_t'=\frac{dc_t}{d\lambda}=\bino{K}{t}(t-K\lambda)\lambda^{t-1}
       (1-\lambda)^{K-t-1} \ \ .
\ee
We again obtain a matrix equation with the same matrix, $\sum_{A'=1}^N
M_{AA'} P'(A')=Z(A)$, where now
\be
  Z(A) = c_A' P(0) + \sum_{r=1}^K \sum_{A'=1}^{N} c_{A-A'+r}' 
         d_{r,A} P(A') \ \ .
\ee
The results quoted in Table~\ref{table1} were obtained by solving
these two matrix equations numerically for $K=3$ and $N=100$ and
applying Eqs.~(\ref{pb}) and (\ref{Cb}).

While this procedure is straightforward, it is not elegant.  We
observe that one can also use Eq.~(\ref{omegaN}) to derive a
$(K-1)$-order differential equation for
\be
  \tilde{\Omega}(z)=\frac{\Omega(z)-\Omega(0)}{z\Omega(0)} \ \ . 
\label{omegatilde}
\ee
The equilibrium distributions of barriers and of the minimum are then 
obtained from
\be
  p(\lambda) = \frac{1}{N} \frac{d}{d\lambda} \Omega'(1)
\label{pdiff}
\ee
and
\be
  C(\lambda) = - \frac{d}{d\lambda} \Omega(0) \ \ .
\label{Cdiff}
\ee
Note that the normalization $\Omega(1)=1$ yields
$\Omega(0)=1/[1+\tilde{\Omega}(1)]$.

Using the definition of Eq.~(\ref{omegatilde}) and the explicit form
of $g_A(z)$, Eq.~(\ref{omegaN}) can be rewritten as
\be
  \frac{z\tilde{\Omega}(z)+1}{f(z)} = 1 + \bino{N-1}{K-1}^{-1}
    \sum_{r=1}^K z^{1-r} \sum_{A>0} \bino{A-1}{r-1} \bino{N-A}{K-r} 
    \frac{P(A)z^{A-1}}{\Omega(0)} \ \ .
\label{diffeq1}
\ee
Replacing $A-1$ by the operator $z\partial_z$ we obtain
\be
  \frac{z\tilde{\Omega}(z)+1}{f(z)} = 1 + \bino{N-1}{K-1}^{-1}
    \sum_{r=1}^K z^{1-r} \bino{z\partial_z}{r-1} 
    \bino{N-1-z\partial_z}{K-r} \tilde{\Omega}(z) \ \ .
\label{diffeq2}
\ee
Here, a binomial coefficient involving an operator $T$ is defined as
\be
  \bino{T}{m} = \frac{T(T-1)\cdots (T-m+1)}{m!} \ \ .
\ee
For example, in the special case of $K=2$ one obtains
\be
  \frac{z\tilde{\Omega}(z)+1}{f(z)} = 1 + \tilde{\Omega}(z) +
  \tilde{\Omega}'(z) \frac{1-z}{N-1} \ \ .
\ee

Eq.~(\ref{diffeq2}) represents an ordinary linear differential
equation for $\tilde{\Omega}(z)$ of order $K-1$.  In particular, it
allows for a systematic expansion in powers of $1/(N-1)$ through an
iterative solution of Eq.~(\ref{diffeq2}).  Such an iterative
procedure is completely algebraic in nature and does not require the
solution of a differential equation.  The quantities $\Omega(z)$,
$p(\lambda)$, and $C(\lambda)$ can then be determined by
Eqs.~(\ref{omegatilde})--(\ref{Cdiff}).  While the matrix equation
above provides explicit expressions for general $K$, it is cumbersome
for large $N$ and $\lambda > 1/K$ where the numerical effort required
to invert the $N\times N$ matrix is appreciable.  The method just
discussed is more suitable for the computation of finite $N$
corrections for any $K$.

\section{Analytical Results in the Thermodynamic Limit}
\label{appendix2}

It is possible to obtain a great many more analytic results in the
thermodynamic limit ($N\rightarrow\infty$).  In this case, the
$d_{r,A}$ of Eq.~(\ref{drA}) are strictly zero for all $r$ except for
$d_{1,A}=1$. Thus, $g_A(z)=z^{-1}$ for all $A>0$.  This simplification
permits the removal of $g_A(z)$ from the sum in Eq.~(\ref{omegaN}),
and we obtain
\be
  \Omega(z)=\Omega(0) f(z) + (\Omega(z)-\Omega(0)) z^{-1} f(z) \ \ .
\ee
Together with the normalization $\Omega(1)=1$ this yields
\be
  \Omega(z)=[1-f'(1)]\frac{(1-z)f(z)}{f(z)-z} 
\ee
with $f(z)$ given after Eq.~(\ref{ct}).

\subsection{Avalanche Distributions}
\label{appendix2.1}

In order to find the probability, $q(t)$, of an avalanche of length
$t$, defined with respect to the threshold $\lambda$, we must look in
detail at the random walks defined by $z^{-1}f(z)$. Let
$\chi(z)=\sum_{t>0} q(t) z^t$ denote the generating function. Further,
define $q_n(t)$ as the probability for having zero active sites for
the first time at the $t$-th time step, given the fact that at time
zero there were $n$ active sites. Denote the generating function for
$q_n(t)$ by $\chi_n(z)$.  (Since an avalanche starts and ends with
zero active sites, $\chi (z) = \chi_0 (z)$.)  The probability
distribution of the number of active sites after the first time step
is the same whether we start with zero or with one active site at time
zero.  It follows that $\chi(z)\equiv \chi_1(z)$.  To compute
$\chi_n$, it is crucial that the random walk defined by $z^{-1}f(z)$
can decrease the number of active sites by at most one. Therefore, if
we start with $n$ active sites, there will be a unique first time when
there are exactly $n-1$ active sites, a unique first time when there
are exactly $n-2$ active sites, and so on. The probability
distribution of these time intervals is precisely given by $\chi_1$,
and therefore
\be 
  \chi_n(z)=\chi(z)^n \ \ . 
\ee
We are now in the position to derive an equation for $\chi$. Suppose
that, at time step zero, there is one active site. Then, the
probability for having $r$ active sites after the first time step
equals the coefficient of $z^r$ in $f(z)$. If there are $r$ active
sites at time zero, then the probability distribution of the first
occurrence of zero active sites is determined by
$\chi_r(z)=\chi(z)^r$.  Combining these facts we see that the
probability distribution for the first occurrence of zero active
sites, having started with zero active sites at time zero, is
$zf(\chi(z))$. The factor of $z$ is due to the fact that we take one
``extra'' time step at the beginning. On the other hand, this quantity
is precisely the avalanche probability distribution. Thus, $\chi$
satisfies
\be 
  \chi=zf(\chi) \ \ .
\ee

In order to find a power series for $\chi$, we need several auxiliary
results.  First, for any integers $A$ and $Y >0$,
\be 
  \sum_{l=1}^Y \bino{Al}{Y-1} \bino{Y}{l} (-1)^l =0 \ \ .
\label{aux1}
\ee
The proof is as follows: the first binomial is the coefficient of
$z^l$ in $z^{(Y-1)/A} (1-z^{1/A})^{-Y}$. The second binomial times
$(-1)^l$ is the coefficient of $z^{Y-l}$ in $(1-z)^Y$.  Thus, the
total sum is the coefficient of $z^Y$ in the product of these two
expressions. The product is
\be 
  \left( 1+ z^{1/A} + z^{2/A} + \cdots + z^{(A-1)/A} \right)^Y
  z^{(Y-1)/A} \ \ ,
\ee
but this does not contain $z^Y$ so that the sum in Eq.~(\ref{aux1})
vanishes. This implies that, for all $\lambda$,
\be
  \sum_{l>0} (1-\lambda)^{(K-1)l+1} \lambda^{l-1} \bino{Kl}{l} 
  \frac{1}{(K-1)l+1}=1 
\ee
by expanding the powers of $(1-\lambda)$ and rewriting the resulting
product of binomials in the form of Eq.~(\ref{aux1}).

Let us now introduce the quantity $y$ defined by
\be
  y \equiv z \lambda (1-\lambda)^{K-1}
\label{y_def}
\ee
and the function $u(y)$ as the solution of
\be
  u(y) [1-u(y)]^{K-1} = y
\label{u_def}
\ee
which has the property that $u(0)=0$.
We claim that 
\be
  \chi=\frac{u(1-\lambda)}{\lambda(1-u)} \ \ .
\label{chisimple}
\ee
Indeed,
\ba
  f(\chi) & = & (1-\lambda+\lambda\chi)^K \nonumber \\
  & = & \frac{(1-\lambda)^K}{(1-u)^K} \nonumber \\
  & = & \frac{(1-\lambda)^K}{y(1-u)/u} \nonumber \\
  & = & \frac{u(1-\lambda)}{z\lambda (1-u)} \nonumber \\
  & = & \frac{\chi}{z} \ \ .
\ea
Now, we are in a position to prove that the power series for $\chi$ is
\be
  \chi(z)=\sum_{l>0} z^l (1-\lambda)^{(K-1)l+1} \lambda^{l-1} 
  \bino{Kl}{l} \frac{1}{(K-1)l+1} \ \ .
\ee
The proof reads as follows:
\ba
  \chi(z) & = & \sum_{l>0} z^l (1-\lambda)^{(K-1)l+1} \lambda^{l-1}
  \bino{Kl}{l} \frac{1}{(K-1)l+1} \nonumber \\ 
  & = & \sum_{l>0} y^l (1-\lambda)\lambda^{-1} \bino{Kl}{l}
  \frac{1}{(K-1)l+1}  \nonumber\\
  & = & (1-\lambda)\lambda^{-1} \sum_{l>0} u^l (1-u)^{(K-1)l}
  \bino{Kl}{l} \frac{1}{(K-1)l+1} \nonumber \\
  & = & \frac{u(1-\lambda)}{\lambda(1-u)} \sum_{l>0} u^{l-1}
  (1-u)^{(K-1)l+1} \bino{Kl}{l} \frac{1}{(K-1)l+1} \nonumber \\ 
  & = & \frac{u(1-\lambda)}{\lambda(1-u)} \ \ . 
\ea

Using this expression, we find for large $t$ that
\be
  q(t) = \frac{1-\lambda}{\lambda}\sqrt{\frac{K}{2\pi (K-1)^3}} \left[
  (1-\lambda)^{K-1}\lambda K^K(K-1)^{-(K-1)} \right]^t t^{-3/2} + {\cal
  O}(t^{-5/2}).
\label{B15}
\ee
The quantity in square brackets is $1$ for the ``critical'' value of
$\lambda = 1/K$.  For this value of $\lambda$, we recover the typical
random walk power law with an exponent of $-3/2$.

\subsection{Temporal Correlations}
\label{appendix2.2}

The final quantity of interest is the temporal correlation function.
Define $S(b,\lambda;\Delta t)$ to be the probability that, given the
fact that at some time a certain site is the minimum with barrier $b$,
it will after $\Delta t$ time steps again be the minimum, for the
first time, with value $\lambda$. Let
$\Theta(b,\lambda;z)=\sum_{\Delta t>0} S(b,\lambda;\Delta t) z^{\Delta
t}$ denote the corresponding generating function.  To evaluate this
quantity we note that the value of the barrier at this particular site
must already be $\lambda$ after the first time step.\footnote{The
possibility of an ``accidental'' change of the barrier at this
particular site (through the other $(K-1)$ randomly selected sites)
has probability zero in the thermodynamic limit and can be neglected.}
After the first time step, a certain number of barriers will have a
value less than $\lambda$. We have to wait until they have all
disappeared before the site in question is again the minimum.  We have
already computed these probabilities, they are just the coefficients
of $\chi_r$ where there are $r$ active sites below the barrier,
$\lambda$. It remains to compute the probability $P_1(b,\lambda;r)$
that, given the fact that the value of the minimum barrier was $b$,
the new barrier value at this site is $\lambda$, and that the number
of active sites with value less than $\lambda$ is $r$ (after the
update).  Then, we have
\be
  \Theta(b,\lambda;z)=z\sum_{r\geq 0} P_1(b,\lambda;r) \chi(z)^r \ \ .
\ee
Let $P_0(b,\lambda;r)$ denote the probability that, given the fact
that at time zero the minimum is $b$, there are $r$ sites with value
less then $\lambda$. We do not need the explicit form of $P_0$.  It is
sufficient to note that, when we integrate over $b$, we will obtain
the probability for having $r$ sites with barriers smaller than
$\lambda$:
\be
  \int_0^1 db \, C(b)\, P_0(b,\lambda;r) = P(r) \ \ .
\ee
From this, we can compute $P_1$. 

Since the minimum is assigned the value $\lambda$, the number of sites
with value smaller than $\lambda$ cannot have changed by more than
$(K-1)$. In addition, we must distinguish between the cases $r=0$ and
$r>0$.  (For $r=0$, $b$ must be larger than $\lambda$.  For $r>0$, $b$
must be smaller than $\lambda$.) This leads to
\ba
  P_1(b,\lambda;r) & = & \bino{K-1}{r} (1-\lambda)^{K-1-r} \lambda^r
                         P_0(b,\lambda;0) \nonumber \\
  & & +\sum_{t>0} \bino{K-1}{r-t+1} (1-\lambda)^{K-1-r+t-1}
      \lambda^{r-t+1} P_0(b,\lambda;t) \ \ .
\ea
This yields
\ba
  \int_0^1 db \, C(b) \, \Theta(b,\lambda;z) & = & z\sum_{r\geq 0}
  \left[\bino{K-1}{r} (1-\lambda)^{K-1-r} \lambda^r P(0)
  \right. \nonumber \\
  & & \qquad \left. +
  \sum_{t>0} \bino{K-1}{r-t+1} (1-\lambda)^{K-1-r+t-1} \lambda^{r-t+1} 
  P(t) \right]\chi(z)^r \nonumber \\
  & = & z[1-\lambda+\lambda \chi(z)]^{K-1} \left[\Omega(0) + 
  \frac{\Omega(\chi(z))-\Omega(0)}{\chi(z)} \right] \nonumber\\
  & = & z[1-\lambda+\lambda \chi(z)]^{K-1} [1-f'(1)] \frac{1-\chi(z)}
  {f(\chi(z))-\chi(z)} \ \ .
\ea
We can further simplify this expression by using the form of $\chi$
given by Eq.~(\ref{chisimple}). We find
\ba
  \Theta(\lambda,z) \equiv \int_0^1 db \,C(b) \, \Theta(b,\lambda;z) 
  & = & z(1-\lambda+\lambda \chi)^{K-1} [1-f'(1)] 
        \frac{1-\chi}{f(\chi)-\chi} \nonumber \\
  & = & \frac{z(1-K\lambda)}{1-\lambda+\lambda\chi} f(\chi) 
        \frac{1-\chi}{f(\chi)-\chi} \nonumber \\
  & = & \frac{z(1-K\lambda)(1-u)}{(1-\lambda)} \frac{\chi}{z}
        \frac{1-\chi}{\chi/z-\chi} \nonumber \\
  & = & \frac{z(1-K\lambda)(1-u)}{(1-z)(1-\lambda)} (1-\chi) \nonumber \\
  & = & \frac{z(1-K\lambda)(1-u)}{(1-z)(1-\lambda)} \left( 
        \frac{\lambda-u}{\lambda(1-u)} \right) \nonumber \\
  & = & \frac{z(1-K\lambda)}{(1-z)(1-\lambda)} \left(
        1-\frac{u}{\lambda}  \right) \ \ .
\label{Omega_lambda_z}
\ea
To obtain a power series for $\Theta$, we need one for $u$. We claim
that
\be  
  u(y)= \sum_{b=1}^{\infty} \frac{(Kb-2)!}{((K-1)b-1)!b!} y^b \ \ .
\ee
We have already demonstrated that 
\be 
  \frac{u}{1-u} = \sum_{l>0} \bino{Kl}{l} \frac{1}{(K-1)l+1} y^l \ \ .
\label{aux2}
\ee
Now,
\ba  
  \sum_{b=1}^{\infty} \frac{(Kb-2)!}{((K-1)b-1)!b!} y^b 
  & = & \frac{K-1}{K}\sum_{b>0} \bino{Kb}{b} \frac{y^b}{Kb-1} \nonumber \\
  & = & \frac{K-1}{K^2} y^{\frac{1}{K}} \int dy \ y^{-1-\frac{1}{K}} 
        \sum_{b>0} \bino{Kb}{b} y^b \ \ ,
\ea
where $\int dy$ stands for the operation $y^r\rightarrow
y^{r+1}/(r+1)$.  From Eq.~(\ref{aux2}) we derive that
\be
  \sum_{b>0}\bino{Kb}{b}y^b=\left[(K-1)y\frac{d}{dy} +1\right] 
  \frac{u}{1-u} \ \ .
\ee
By differentiating $u(1-u)^{K-1}=y$ with respect to $y$ we find
\be 
  u'(y)=\frac{u(y)(1-u(y))}{y(1-Ku(y))} \ \ .
\ee
Using this result, we find that
\be
  \sum_{b>0} \bino{Kb}{b} y^b  = \frac{Ku}{1-Ku}
\ee
and
\be
  \frac{d}{dy} \left( y^{-\frac{1}{K}} u \right) = y^{-1-\frac{1}{K}}
  \frac{(K-1)u}{K(1-Ku)} \ \ .
\ee
Armed with this we obtain 
\ba 
  \sum_{b=1}^{\infty} \frac{(Kb-2)!}{((K-1)b-1)!b!} y^b 
  & = & \frac{K-1}{K^2} y^{\frac{1}{K}} \int dy \ y^{-1-\frac{1}{K}} 
        \sum_{b>0} \bino{Kb}{b} y^b \nonumber \\
  & = & \frac{K-1}{K} y^{\frac{1}{K}} \int dy \ y^{-1-\frac{1}{K}} 
         \frac{u}{1-Ku} \nonumber \\
  & = & y^{\frac{1}{K}} \int dy \ \frac{d}{dy} \left( y^{-\frac{1}{K}}
        u \right) \nonumber \\
  & = & u \ \ .
\ea
This proves the claimed power series for $u$.  

We have now obtained the result  
\be
  \Theta(\lambda;z)=\frac{z(1-K\lambda)}{(1-z)(1-\lambda)}\left\{
  1-\frac{1}{\lambda} \sum_{b=1}^{\infty} \frac{(Kb-2)!}{((K-1)b-1)!
  b!} \left[ z\lambda (1-\lambda)^{K-1} \right]^b \right\} \ \ .
\ee
The next step is to integrate over $\lambda$. If $\lambda> 1/K$, the
site will never again become the minimum.  Thus, we need only
integrate from $0$ to $1/K$. The integrals involved are easy, since
\be
  (1-K\lambda)\lambda^{b-1} (1-\lambda)^{(K-1)b-1} = 
  \frac{1}{b} \frac{d}{d\lambda} \left[\lambda^b (1-\lambda)^{(K-1)b} 
  \right] \ \ .
\ee
The final result is
\ba
  \Theta(z)\equiv\int^{1/K}_0 d\lambda\, \Theta(\lambda;z) & = & 
  \frac{z}{1-z} \left\{ 1-(K-1) \ln \left(\frac{K}{K-1} \right) 
 \right. \nonumber \\
& & \qquad  \qquad \left.
  - \sum_{b=1}^{\infty} \frac{(Kb-2)!}{b((K-1)b-1)!b!} \left[ 
  \frac{z(K-1)^{K-1}}{K^{K}} \right]^b \right\} \ \ .
\label{aux3}
\ea
Unfortunately, it is not easy to extract the behaviour of the
coefficients for large powers of $z$ from this expression due to the
factor $1/(1-z)$.  This behaviour can be obtained more easily from a
power series constructed about $z=1$ instead of $z=0$.  The desired
series can be derived from Eq.~(\ref{aux3}) by writing it as
\be
  \frac{d}{dz} \left( \frac{(1-z)\Theta(z)}{z} \right) = 
  - \frac{\tilde{u}(z)}{z} \ \ ,
\ee
where $\tilde{u}(z)$ is the solution of
\be 
  \tilde{u}(1-\tilde{u})^{K-1} = z\frac{(K-1)^{K-1}}{K^{K}} \ \ .
\label{aux4}
\ee
which satisfies $\tilde{u}(0)=0$. The function $\tilde{u}$ has a
series expansion
\be 
  \tilde{u}=\frac{z}{K} \left( 1+\sum_{j>0}a_j(1-z)^{j/2}\right) \ \ ,
\label{aux6}
\ee
and from Eq.~(\ref{aux4}) one can easily determine a few of the
coefficients $a_j$.  That this is a series in $\sqrt{1-z}$ rather than
$(1-z)$ is due to the fact that the function
$f(\tilde{u})=\tilde{u}(1-\tilde{u})^{K-1}$ has a maximum at
$\tilde{u}=1/K$.  Given the series Eq.~(\ref{aux6}), $\Theta(z)$ is
recovered from
\be
  \Theta(z)=\frac{z}{1-z} \int_z^1 dy \, \frac{\tilde{u}(y)}{y} \ \ ,
\ee
where the proper integration limits follow from the requirement that
$\Theta(z)$ should have no singularity as $z\rightarrow 1$. The first
few terms in the expansion of $\Theta(z)$ are
\begin{eqnarray} 
  \Theta(z) & = & 
  \frac{z}{K} \left\{1-\frac{2}{3} \sqrt{\frac{2(K-1)}{K}} (1-z)^{1/2}
  + \frac{5K-4}{6K} (1-z) \right. \nonumber \\
  & & \qquad -\frac{47K^2-71K+26}{90K(K-1)}\sqrt{\frac{2(K-1)}{K}}
      (1-z)^{3/2} \nonumber \\ 
  & & \qquad +\frac{2(134 K^3-291 K^2+204 K-46)}{405 K^2 (K-1)} 
      (1-z)^2 \nonumber \\
  & & \qquad  - \frac{6409 K^4-17954 K^3+18261 K^2-7964 K+1252}
  {15120 K^2 (K-1)^2} \sqrt{\frac{2(K-1)}{K}} (1-z)^{5/2}  \nonumber \\ 
  & & \qquad \left.  + {\cal O}((1-z)^3) \right\} \ \ .
\label{aux5}
\end{eqnarray}
Note that $\Theta(1)=1/K$ as expected.  Thus, there is a probability
of $(K-1)/K$ that a site ``dies'' upon updating and will never become
active again.

For $K=2$, there exists a simple closed expression for $\Theta(z)$,
\be
  \Theta(z)=\frac{z}{1-z} \left\{ \sqrt{1-z}-\ln (1+\sqrt{1-z})
  \right\} \ \ .
\ee
The series expansion, Eq.~(\ref{aux5}), for $\Theta(z)$ around $z=1$
determines the asymptotic behaviour of $S(\Delta t)$ for large $\Delta
t$, and we find once more a power law with the (leading) random walk
exponent $-3/2$,
\ba
  S(\Delta t)&=&\sqrt{\frac{2(K-1)}{\pi K}} \left\{ \frac{1}{3K} 
  (\Delta t)^{-3/2}+\frac{14K^2-2K-13}{60 K^2 (K-1)} (\Delta t)^{-5/2}
  \right. \nonumber \\
  & &  \qquad\qquad\qquad + \frac{169K^4+142K^3-543K^2-80K+313}
  {2016 K^3(K-1)^2} (\Delta t)^{-7/2} \nonumber \\ 
  & & \qquad \qquad \qquad \left. + {\cal O}((\Delta t)^{-9/2}) 
  \right\} \ \ .
\label{SDeltat}
\ea
These are the ``first-return'' probabilities of Sec.~\ref{section4}.
It is far more difficult to compute the desired temporal correlation
function $P_t(\Delta t)$, i.e., the ``all-return'' probabilities.  The
na\"{\i}ve expectation would be to generalize Eq.~(\ref{Q_from_P}),
leading to Eq.~(\ref{approx_rel}).  This is not correct, however.  The
underlying assumption is that the ensemble of configurations at the
moment of a first return (after a fixed number of steps) is equal to
the ensemble of all configurations.  One can show numerically that
this assumption is not satisfied.  This implies that, in order to
perform an exact calculation of the all-return probabilities, one
needs the simultaneous probability distribution of all barriers below
a given threshold at all times.  Fortunately, substitution of $z=1$ in
Eq.~(\ref{Q_from_P}) corresponds to averaging over all ensembles of
configurations at the moment of first returns.  This means that the
previous objections do not apply in this situation and that
Eq.~(\ref{inverse}) is, indeed, exact.  This is important for the
discussion in Sec.~\ref{section4}.

One can derive an improved equation for the all-return probabilities
by taking into account some of the correlations that have been
neglected in Eq.~(\ref{approx_rel}).  Specifically, we can account for
the different probability distribution of the minimum in the two
ensembles mentioned previously (at the moment of a first return and
averaged over all configurations).  The improved generating function
for the all-return probabilities, $\Phi(z)=\sum_{\Delta t>0}
P_t(\Delta t) z^{\Delta t}$, can thus be written as 
\be 
  \Phi(z) =
  \sum_{n\ge 1} \int_0^{1} d\lambda_{1} \, C(\lambda_1) \, \prod_{i=1}^n
  \int_0^{1/K} d\lambda_{i+1} \ \Theta(\lambda_i,\lambda_{i+1};z) \ \ .
\ee 
Note that this generating function can also be obtained from 
\be
  \Phi(z) = \int_0^1 \, d\alpha \, C(\alpha)\, \int_0^{1/K} d\beta \
  \Phi(\alpha,\beta;z) \ \ , 
\ee 
where $\Phi(\alpha,\beta;z)$ is given by the integral equation 
\be 
  \Phi(\alpha,\beta;z) =\Theta(\alpha,\beta;z) + \int_0^{1/K} d\lambda \
  \Theta(\alpha,\lambda;z) \Phi(\lambda,\beta;z) \ \ .
\label{int_eq}
\ee
Eq.~(\ref{int_eq}) does not appear to admit explicit analytic
solution.  Thus, we have elected to use the approximate relationship,
Eq.~(\ref{inverse}), to obtain the asymptotic behaviour of the
all-return probabilities.  While it is not simple to invert
$1-\Theta(z)$ in closed form, we can again employ Eq.~(\ref{aux5}) to
extract the asymptotic behaviour of $P_t(\Delta t)$ for large $\Delta
t$.  After some algebraic effort we obtain
\ba
  P_t(\Delta t)&=&\sqrt{\frac{2(K-1)}{\pi K}} \left\{ \frac{K}{3(K-1)^2}
  (\Delta t)^{-3/2}+\frac{42K^2+24K+1}{180(K-1)^3} (\Delta t)^{-5/2}
  \right.\nonumber \\
  & &  \qquad\qquad\qquad + \frac{1521K^4+802K^3-3417K^2+36K+17}
  {18144K(K-1)^4} (\Delta t)^{-7/2}
  \nonumber \\  & & \qquad \qquad \qquad \left.
  +{\cal O}((\Delta t)^{-9/2}) \right\}
\label{allreturn}
\ea
with the now-familiar leading exponent $-3/2$.  This approximate
solution for the all-return probabilities does exceedingly well in
describing the data as can be seen from Fig.~\ref{figPtrandom}.

\newpage

\begin{figure}
\caption{Simulation results for the spatial correlation function of
the minimum as defined in Sec.~\protect\ref{section4} for the nearest
neighbour model ($N_{\rm sites}=500$, $10^8$ updates). }
\label{figPxlocal}
\end{figure}

\begin{figure}
\caption{Simulation results for the temporal correlation function of
the minimum as defined in Sec.~\protect\ref{section4} for the nearest
neighbour model ($N_{\rm sites}=500$, $10^8$ updates). Crosses
indicate first-returns, diamonds all-returns.}
\label{figPtlocal}
\end{figure}

\begin{figure}
\caption{Simulation results for the spatial correlation function of
the minimum for the random neighbour model with $K=3$ ($N_{\rm sites}=500$, 
$10^5$ updates). }
\label{figPxrandom}
\end{figure}

\begin{figure}
\caption{Simulation results for the temporal correlation function of
the minimum for the random neighbour model with $K=3$ ($N_{\rm sites}=100,000$,
$2\times 10^7$ updates). Crosses indicate first-returns, diamonds
all-returns.  Also shown are the predictions of
Eqs.~(\protect\ref{SDeltat}) (solid line) and
(\protect\ref{allreturn}) (dashed line).}
\label{figPtrandom}
\end{figure}

\newpage

\begin{table}
\caption{Distribution of barrier heights, $p(\lambda)$, and of the
minimum, $C(\lambda)$, for $N=100$ sites and $K=3$ as a function of
$\lambda$.  Results are shown for the mean field approximation, the
(exact) active site model, and the simulation of $10^8$ updates.  The
inadequacies of mean field theory are especially visible in the
transition region around $\lambda=1/3$.}
\label{table1}
\begin{tabular}{ccccccc}
$\lambda$ & \multicolumn{3}{c}{$p(\lambda)$} & 
\multicolumn{3}{c}{$C(\lambda)$} \\ 
{} & {\sc mean field} & {\sc active sites} & {\sc data} & {\sc mean
field} & {\sc active sites} & {\sc data} \\ \tableline
0.00 & 0.030000 & 0.030000 & 0.029951 & 3.000000 & 3.000000 & 2.996709 \\
0.05 & 0.035103 & 0.033735 & 0.033757 & 2.989478 & 2.992299 & 2.994195 \\
0.10 & 0.042277 & 0.039979 & 0.039978 & 2.974686 & 2.979425 & 2.980276 \\
0.15 & 0.053083 & 0.051466 & 0.051507 & 2.952407 & 2.955741 & 2.956010 \\
0.20 & 0.071137 & 0.075558 & 0.075558 & 2.915182 & 2.906067 & 2.907232 \\
0.25 & 0.107012 & 0.135857 & 0.135955 & 2.841212 & 2.781739 & 2.782052 \\
0.30 & 0.207912 & 0.318852 & 0.318660 & 2.633172 & 2.404429 & 2.404860 \\
0.31 & 0.252919 & 0.388840 & 0.389023 & 2.540372 & 2.260124 & 2.261542 \\
0.32 & 0.319176 & 0.475928 & 0.475627 & 2.403760 & 2.080561 & 2.081183 \\
0.33 & 0.422562 & 0.581861 & 0.581703 & 2.190595 & 1.862142 & 1.863407 \\
0.34 & 0.592244 & 0.706248 & 0.705545 & 1.840734 & 1.605675 & 1.604468 \\
0.35 & 0.863716 & 0.845113 & 0.845008 & 1.280998 & 1.319354 & 1.320481 \\
0.36 & 1.190983 & 0.990031 & 0.989369 & 0.606221 & 1.020556 & 1.020026 \\
0.37 & 1.397891 & 1.128966 & 1.128950 & 0.179607 & 0.734090 & 0.734467 \\
0.38 & 1.465339 & 1.249437 & 1.247942 & 0.040538 & 0.485697 & 0.485071 \\
0.39 & 1.480985 & 1.342799 & 1.342416 & 0.008279 & 0.293197 & 0.292592 \\
0.40 & 1.484215 & 1.407036 & 1.408536 & 0.001619 & 0.160750 & 0.160894 \\
0.45 & 1.485000 & 1.484061 & 1.483476 & 0.000000 & 0.001937 & 0.002008 \\
0.50 & 1.485000 & 1.484999 & 1.484494 & 0.000000 & 0.000003 & 0.000002 \\
0.55 & 1.485000 & 1.485000 & 1.486311 & 0.000000 & 0.000000 & 0.000000 \\
\end{tabular}                                                        
\end{table}                                                          
                                                                     
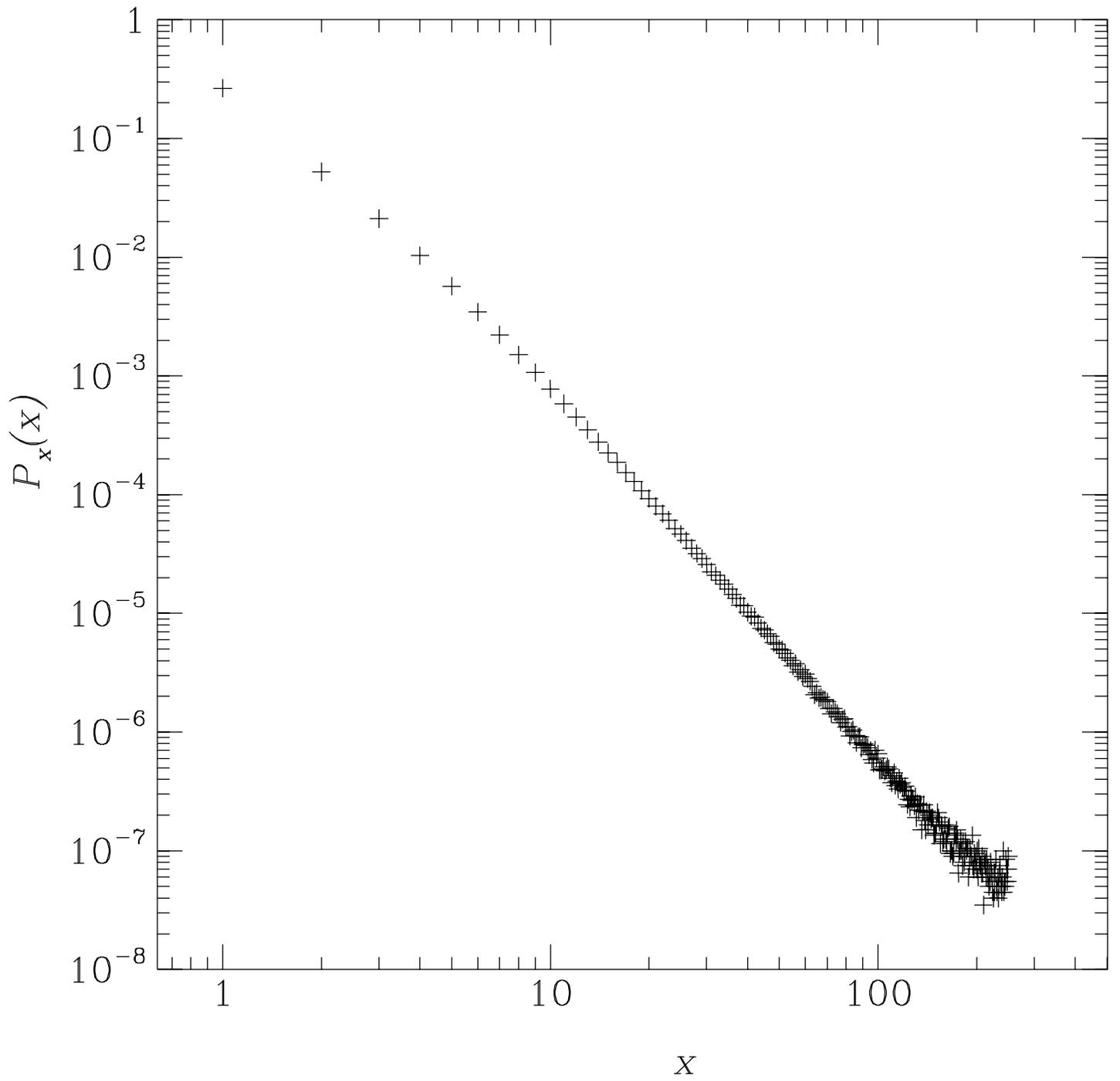
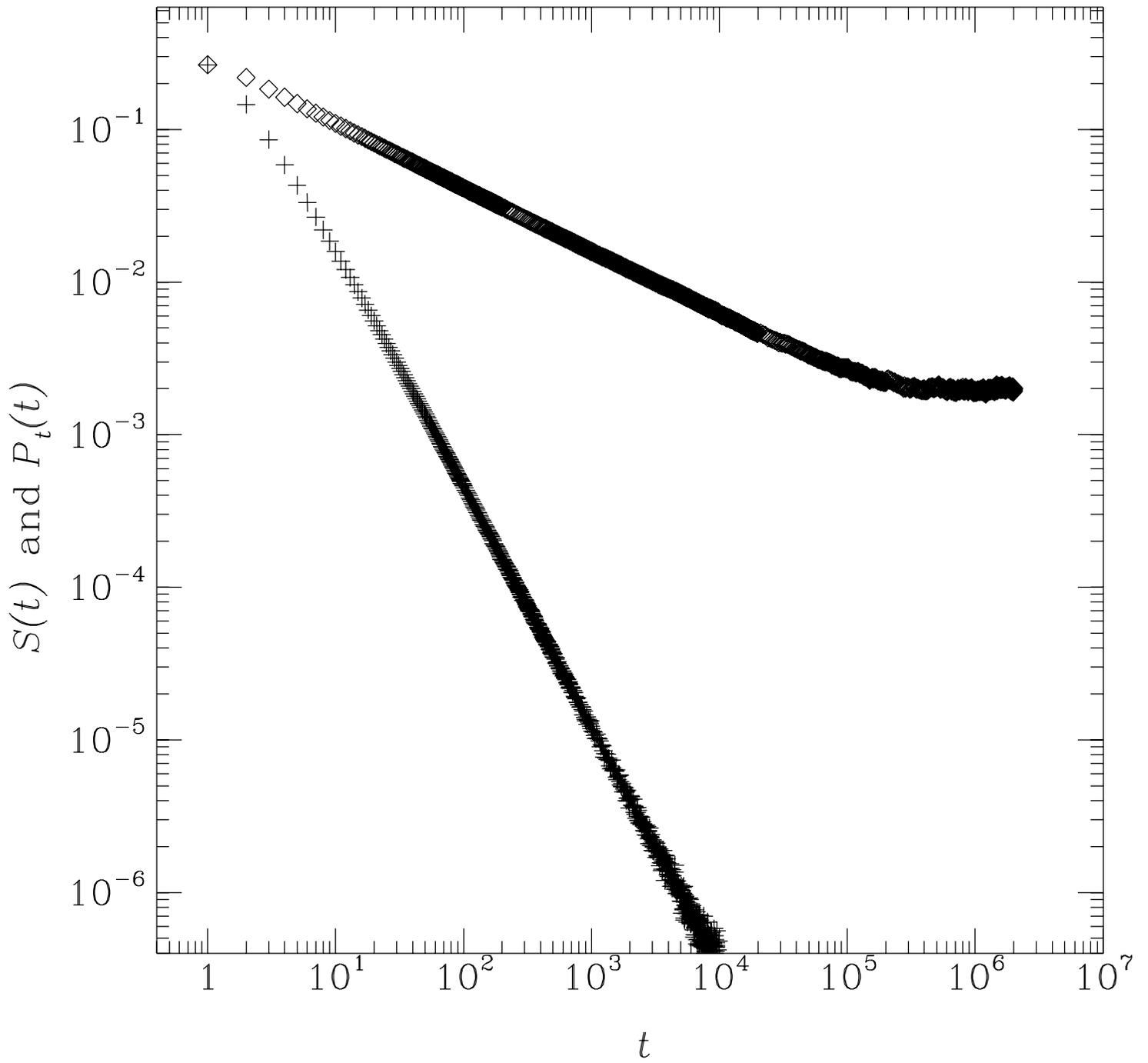
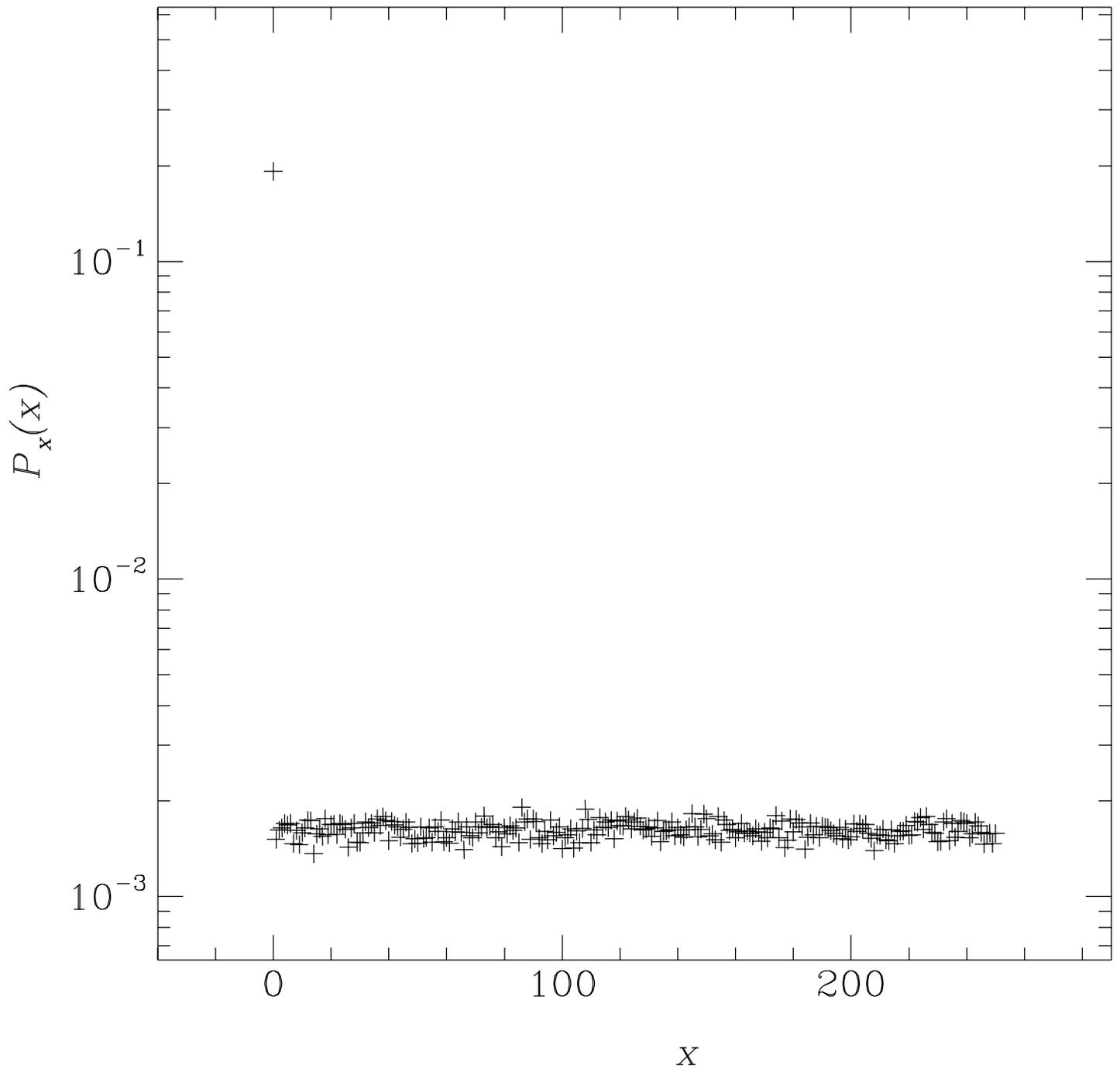
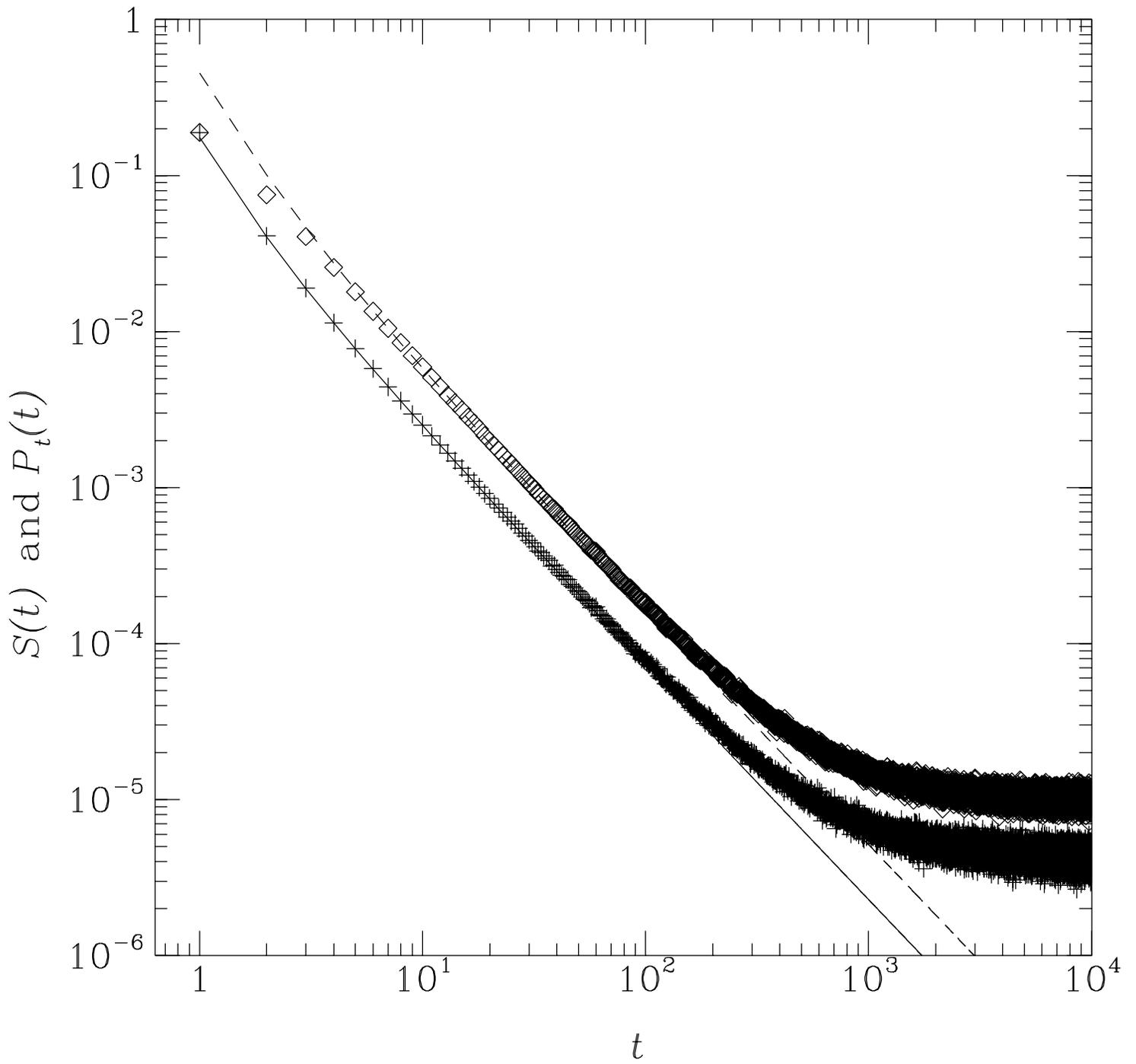
\end{document}